\documentclass[aps,twocolumn,secnumarabic,nobalancelastpage,amsmath,amssymb,
nofootinbib]{revtex4}

\usepackage{graphics}      
\usepackage{graphicx}      
\usepackage{longtable}     
\usepackage{url}           
\usepackage{bm}            

\usepackage[english]{babel}
\usepackage[utf8x]{inputenc}
\usepackage[T1]{fontenc}

\usepackage[a4paper,top=2cm,bottom=2cm,left=2cm,right=2cm,marginparwidth=1.75cm]{geometry}

\usepackage{amsmath}
\usepackage{graphicx}
\usepackage[colorinlistoftodos]{todonotes}
\usepackage[colorlinks=true, allcolors=blue]{hyperref}
\usepackage{booktabs}

\usepackage{setspace}

\usepackage{amsmath}

\usepackage{graphicx}

\begin{document}


\title{An Experimental and Theoretical Insights into the Dielectric Properties of (Li, Nd) Co-doped ZnO Ceramics} 
\author         {Md. Zahidur Rahaman\textsuperscript{\S \textdagger}, Hidekazu Tanaka\textsuperscript{\textdaggerdbl}, A.K.M. Akther Hossain\textsuperscript{\textdagger}}

\thanks{Corresponding author. \small \textsuperscript{\S \textdagger}\textit{\href{mailto:zahidur.physics@gmail.com}{zahidur.physics@gmail.com}, \href{mailto:md_zahidur.rahaman@unsw.edu.au}{md\_zahidur.rahaman@unsw.edu.au}} (M. Z. Rahaman); \small\textsuperscript{\textdagger}\textit{\href{mailto: akmhossain@phy.buet.ac.bd}{akmhossain@phy.buet.ac.bd}} (A.K.M. Akther Hossain).}

\affiliation    {\textsuperscript{\S}School of Materials Science and Engineering, University of New South Wales, Sydney 2052, Australia\\\textsuperscript{\textdagger}Department of Physics, Bangladesh University of Engineering and Technology, Dhaka-1000, Bangladesh\\ \textsuperscript{\textdaggerdbl}Institute of Scientific and Industrial Research, Osaka University, Ibaraki, Osaka 567-0047, Japan}
\date{\today}


\begin{abstract}
\section*{Abstract} 
In this work, we report the combined effect of donor (Nd) and acceptor (Li) co-doping at the Zn-site of ZnO ceramics on structural, microstructural and dielectric properties. Combining experimental observations with DFT based theoretical study, we have shown that before experimental fabrication DFT based first principles study can be used as a good indication to have prior qualitative assessment of a dielectric medium. For implementing this objective various Li and Nd co-doped ZnO ceramics have been synthesized through the conventional solid-state reaction route. Quantitative XRD analysis reveals the formation of wurtzite hexagonal structured ZnO having space group \textit{P6\textsubscript{3}mc}. Meanwhile, FESEM micrographs confirm the formation of randomly aligned non-uniform grains in size and shape. We show that the average grain size distribution and density of the studied compositions are two tuning factors to control the dielectric properties of these compounds. Though the value of dielectric constant is decreased with the increase in doping content, the optimum composition Nd\textsubscript{0.005}Li\textsubscript{0.005}Zn\textsubscript{0.99}O exhibits slightly lower dielectric constant (\(\approx\) 2066 at 1 KHz) than pristine ZnO but relatively very low dielectric loss (\(\approx\) 0.20 at 1 KHz) at room temperature than pure ZnO ceramics sintered at 1623 K. For understanding the dielectric relaxation mechanism in the studied ceramics, complex impedance spectra analysis have also been performed and discussed thoroughly. This study provides a new insight for further development of ceramic materials with improved dielectric properties.

\textbf{Keywords:} ZnO ceramic, Dielectric constant, Dielectric loss, DFT study.

\end{abstract}

\maketitle

\section*{I. INTRODUCTION}

At present, dielectric materials with colossal dielectric constant (CDC)
have created great interest in the scientific community due to their
potential applications as high energy density storage such as the
ceramic power capacitors and the multi-layer ceramic capacitors (MLCC)
\cite{1}. An ideal dielectric material should have a number of
characteristics including high-frequency stability, high temperature
stability, colossal permittivity (CP) and considerably low dielectric
loss \cite{2}. However, it is still very difficult to maintain all of the
above requirements in a single material. The giant dielectric property
has been reported in rare-earth (RE) and transition metal doped
BaTiO\textsubscript{3} \cite{3}, NiO \cite{4}, CuO \cite{5},
TiO\textsubscript{2} \cite{6}, and ZnO \cite{1, 7} with relative
permittivity \(\geq\)10\textsuperscript{3}. Among the inorganic ceramics
various metal doped BaTiO\textsubscript{3} is regarded as one of the
most suitable candidates for the application in ferroelectric
capacitors. However, BaTiO\textsubscript{3} ceramic exhibits the CP
behavior for a narrow temperature range close to the temperature of the
phase transition from ferroelectric to paraelectric phase \cite{8}.
The CaCu\textsubscript{3}Ti\textsubscript{4}O\textsubscript{12} (CCTO) is
considered as another promising CP material having the temperature
independent dielectric constant of about 10\textsuperscript{5} \cite{9}.
Unfortunately, poor frequency stability and considerably high dielectric
loss (\textgreater{}0.2) makes CCTO less suitable for the proposed
technological applications. Though relaxor ferroelectric materials
exhibit CP behavior with relatively low dielectric loss, it is still
constrained in a narrow temperature range (even close to the temperature
of phase transition) \cite{10}. Therefore, a lot of efforts should be
given to find novel CP materials with a very wide range of frequency and
temperature stability and sufficiently low dielectric loss.

Wurtzite hexagonal type ZnO is a well known wide band gap semiconductor
(direct band gap 3.3 eV) in the transition metal oxide family. ZnO is
commercially available cheap metal oxide and it is comparatively easier
to prepare the pure form of ZnO. Due to this reason to find a true ZnO
based CP material will be a great leap in forward. A lot of efforts have
already been given to make pure ZnO a true CP candidate.
Tripathi~\emph{et al.} report achieving high dielectric constant of
10\textsuperscript{4} at low-frequency region in ZnO nanostructure
prepared by soft chemical approach \cite{11}. However, the prepared
nanostructure is not sintered and with the increase in frequency, the
dielectric constant is rapidly decreased. Room temperature CP behavior
was also observed in high pressure treated ZnO ceramic by Xuhai \emph{et
al.} in 2014 \cite{12}. Grain size engineering and chemical modification
through doping can be another possible way to improve the dielectric
properties of ZnO ceramics. Recently, Dong Huang \emph{et al.} report
having giant dielectric constant up to 3800 with low dielectric loss of
0.11 at 1 KHz in (Li, In) co-doped ZnO ceramic \cite{1}. They suggest
that oxygen defects can be the possible origin of CP behavior in (Li,
In) co-doped ZnO. CP behavior is also observed in (Li, Fe) co-doped ZnO
by You \emph{et al.} due to the formation of defect dipole by acceptor
Li and donor Fe atom \cite{7}. However, (Li, Fe) co-doped ZnO shows poor
frequency stability.

In the present study, (Li, Nd) co-doped ZnO ceramics have been prepared
for the first time to explore the dielectric behavior through basic
study. The structural, morphological and dielectric properties of Li
(acceptor) and Nd (donor) co-doped ZnO ceramics have been investigated
through various experimental techniques. For further justification and
interpretation of the obtained experimental results a DFT based first
principles study has been carried out by using the plane wave
pseudopotential approach.

\section*{II. METHOD}
\subsection*{A. Experimental}

The conventional solid-state reaction technique is used for synthesizing
the (Li, Nd) co-doped ZnO ceramics with formula
(Nd\textsubscript{0.5}Li\textsubscript{0.5})\textsubscript{x}Zn\textsubscript{1-x}O where, x = 0.00, 0.01, 0.03, 0.05, and 0.10. ZnO (≥ 99.9\%),
Li\textsubscript{2}CO\textsubscript{3} (≥ 99.9\%) and
Nd\textsubscript{2}O\textsubscript{3} (≥ 99.9\%) are taken as the starting materials (raw materials). Stoichiometric amount of the raw
materials (powder form) are properly mixed in an agate mortar by hand
milling for about 6 hours. Acetone (Propanone) is used as a volatile
organic liquid for making the mixer more homogeneous. During the mixing
and grinding process acetone volatilizes gradually and evaporates
completely after 10 to 15 minutes. After grinding and mixing properly,
the dried powders are then calcined in an alumina crucible by using a
programmable electric furnace at 1073 K for 5 hours in
the air with a heating and cooling rate of about 10 and 5
\textsuperscript{o}C/min, respectively. The calcined powder is then
re-milled for 4 hours for ensuring proper homogenization. Finally, the
dried fine powders are used to prepare the disc shaped pellet (diameter 12-13 mm and thickness 1-1.5 mm) by applying
uniaxial pressure of 4000 psi for 1 min through a hydraulic press. A small drop of poly vinyl alcohol (PVA) is mixed as a binder for preparing each of the green body. The green samples are then sintered at 1473, 1523, 1573, 1623 and 1648 K in the air for burning out the PVA and densification. The heating and cooling rate in this case is same as the calcination process. The sintered samples are then polished to remove roughness of the surface and any oxide layer formed during the sintering process.

The crystal structure and phase purity of the studied compositions are investigated by using an advanced X-ray diffractometer (Model-Philips PANalytical X’PERT-PRO, Cu-K\(\alpha\) is used as target with incident wavelength, \textit{\(\lambda\)} = 1.540598 \AA). The microstructural analysis of the surface of sintered samples is performed by using the field emission scanning electron microscope (FESEM, JEOL, JSM-7600F). The average grain size is evaluated by using the linear intercept
technique through the relation, \(\overline{D} = 1.56\overline{L}\)
\cite{22}; where \(\overline{D}\) stands for average grain size, and
\(\overline{L}\) is the average intercept length over a large number of
grains as measured on the plane of the sample. The experimental or bulk density of the selected compositions is calculated by using the following relation, 
\begin{equation}
   \rho_{exp} = \frac{M}{\pi r^2 h}
\label{eq:first-equation}
\end{equation}
where \emph{r} is the radius of the pellet shaped sample and \emph{h} is the thickness of the pellet. The theoretical density of all the pellet shaped specimens is evaluated by using the following relation,
\begin{equation}
   \rho_{th} = \frac{ZM}{N_{A} V}
\label{eq:first-equation}
\end{equation}
where \emph{Z} denotes the number of formula unit per unit cell,
\emph{M} is the molecular weight, \emph{N\textsubscript{A}} is defined
as the Avogadro's number (6.023 \(\times\) 10\textsuperscript{23}
/mole), and \emph{V} is the volume of the unit cell. The porosity of all the selected compositions is evaluated by using the following expression,
\begin{equation}
   P (\%) = \left(\frac{\rho_{th} - \rho_{B}}{\rho_{th}}\right) × 100
\label{eq:first-equation}
\end{equation}
where \(\rho\)\textsubscript{th} and \(\rho\)\textsubscript{B} is defined as the theoretical and bulk (experimental) density, respectively. The measurements of dielectric properties are performed by using a precision Impedance Analyzer (Wayne Kerr Impedance Analyzer, 6500B). For dielectric measurement, the pellet shaped samples are first polished to remove roughness of the surface and contamination of any other oxides on the surface during the sintering process. Both sides of the samples are then painted with conducting silver paste for ensuring good electrical contact.  The ac-conductivity of all the sintered samples is evaluated from the dielectric constant data by using the following expression \cite{13},  
\begin{equation}
   \sigma_{ac} = \varepsilon^\prime \varepsilon_{o} \omega tan\delta
\label{eq:first-equation}
\end{equation}
where \(\omega\) (= \(2\pi f\)) defines the angular frequency. A precision Impedance Analyzer (Wayne Kerr Impedance Analyzer, 6500B) is used for the measurement of real (\(Z^{\prime}\)) and imaginary (\(Z^{\prime\prime}\)) part of the complex impedance as a function of frequency at room temperature. The real and imaginary part of the electric modulus are obtained from the impedance data according to the following relations \cite{14},  
\begin{equation}
   M^\prime = \frac{\varepsilon^\prime}{(\varepsilon^{\prime2} + \varepsilon^{\prime\prime2})} = \omega C_{o}Z^{\prime\prime} 
\label{eq:first-equation}
\end{equation}
\begin{equation}
   M^{\prime\prime} = \frac{\varepsilon^{\prime\prime}}{(\varepsilon^{\prime2} + \varepsilon^{\prime\prime2})} = \omega C_{o}Z^\prime
\label{eq:first-equation}
\end{equation}

\subsection*{B. Theoretical}

The Density Functional Theory (DFT) dependent plane wave pseudopotential approach executed within the CASTEP code is used to carry out the theoretical calculations \cite{15, 16}. Generalized Gradient Approximation (GGA) with the Perdew, Burke and Ernzerhof (PBE) exchange correlation functional is used for estimating the exchange correlation energy \cite{17}. The plane wave cutoff energy of 380 eV is used to expand the wave function. The Monkhorst-Pack scheme \cite{20} is used to comprise the K-point sampling of the Brillouin zone. The geometry optimization is performed by using 4 × 4 × 2 k-points. The number of k-points is enough for good convergence of the pure as well as doped ZnO. Vanderbilt type ultrasoft pseudopotential is used for describing the electron ion interaction \cite{18}. The geometry optimization of both the pristine and doped system is executed by the Broyden-Fletcher-Goldfarb-Shanno (BFGS) relaxation scheme \cite{19}. The dielectric properties of the selected compositions are evaluated by using the CASTEP tool based on the standard DFT Kohn-Sham orbitals \cite{21}.

\section*{III. RESULTS AND DISCUSSION}
\subsection*{A. Structural Analysis}
\begin{figure}[h]
\includegraphics[width=7.2cm]{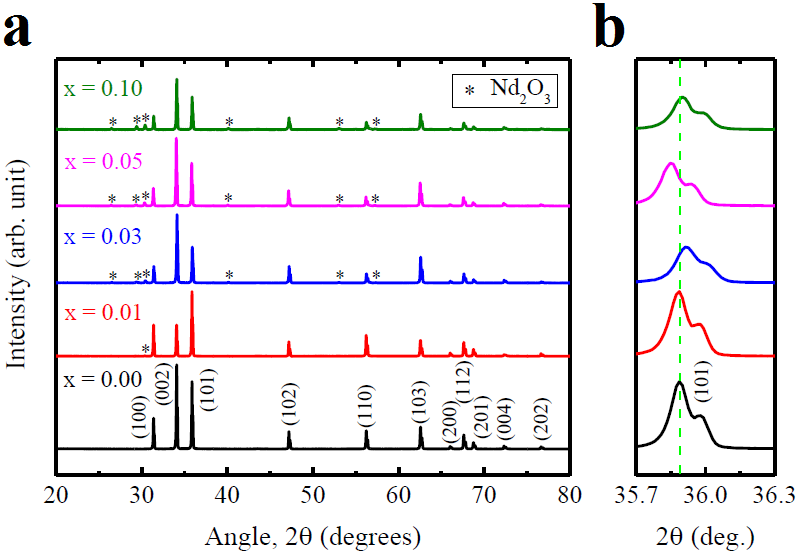}
\caption{(a) The X-ray diffraction pattern of (Nd\textsubscript{0.5}Li\textsubscript{0.5})\textsubscript{x}Zn\textsubscript{1-x}O ceramics sintered at 1623 K and (b) the shifting of peak corresponding to (101) plane.\label{fig:Fig. 1}}
\end{figure}

The structural properties of (Nd\textsubscript{0.5}Li\textsubscript{0.5})\textsubscript{x}Zn\textsubscript{1-x}O (x = 0.00, 0.01, 0.03, 0.05, and 0.10) ceramics are analyzed thoroughly by using X-Ray Diffraction (XRD) techniques. The XRD patterns of the (Nd\textsubscript{0.5}Li\textsubscript{0.5})\textsubscript{x}Zn\textsubscript{1-x}O ceramics sintered at 1623 K are illustrated in Fig. \ref{fig:Fig. 1}. 
The diffraction peaks of (Li, Nd) co-doped ZnO can be indexed to the wurtzite hexagonal type zinc oxide [JCPDS 89-7102 (ZnO)] having space group \textit{P6\textsubscript{3}mc} (186). In the pristine ZnO bulk material no secondary phases are appeared. However, the XRD patterns of (Li, Nd) co-doped ZnO contain some secondary peaks. The XRD pattern of 1\% co-doped sample contains only one impurity peak and with the increase in dopant contents the number of secondary peaks are increased. The impurity peaks are appeared due to the presence of Nd\textsubscript{2}O\textsubscript{3} phase (JCPDS 83-1346) with hexagonal type structure having space group \textit{P6\textsubscript{3}/mmc} (194) in the desired compositions. It has also been reported previously that Nd doped ZnO contains some extra peaks of Nd\textsubscript{2}O\textsubscript{3} phase \cite{23}. The presence of Nd\textsubscript{2}O\textsubscript{3} phase in the desired compositions implies that Neodymium (Nd) atoms did not insert completely into the crystalline lattice site resulting inhomogeneous solid solution. However, the shift in peak corresponding to (101) plane as shown in Fig. \ref{fig:Fig. 1}(b) confirms the partial incorporation of Li and Nd at the Zn site of ZnO ceramic. The reason behind the partial incorporation of Li and Nd at the Zn site can be attributed to the comparatively large ionic radius of Nd compared to Zn. The ionic radius of Zn\textsuperscript{2+} is about 0.074 nm and Nd\textsuperscript{3+} is 0.098 nm. Due to this large ionic radius Nd\textsuperscript{3+} encounters some difficulty to replace the Zn\textsuperscript{2+} site of ZnO bulk ceramic. Generally, the incorporation of Li at Zn site shifts the peak in the higher angle (implying the decrease in lattice parameters as a result of decreasing the interplaner spacing between the lattices) and incorporation of Nd shifts the peak in the lower angle (implying the increase in lattice parameters) \cite{24, 25}. From Fig. \ref{fig:Fig. 1}(b) it is evident that the shift in peak corresponding to (101) plane is not uniform. The presence of secondary phase and partial incorporation of dopants into the lattice site can be one reason behind these anomalous shift in peaks with composition. However, shifting in peaks toward higher angle results the enhancement of lattice strain and the strain is minimum when peaks shift toward the lower angle \cite{23}.

\begin{figure}[ht]
\includegraphics[width=8cm]{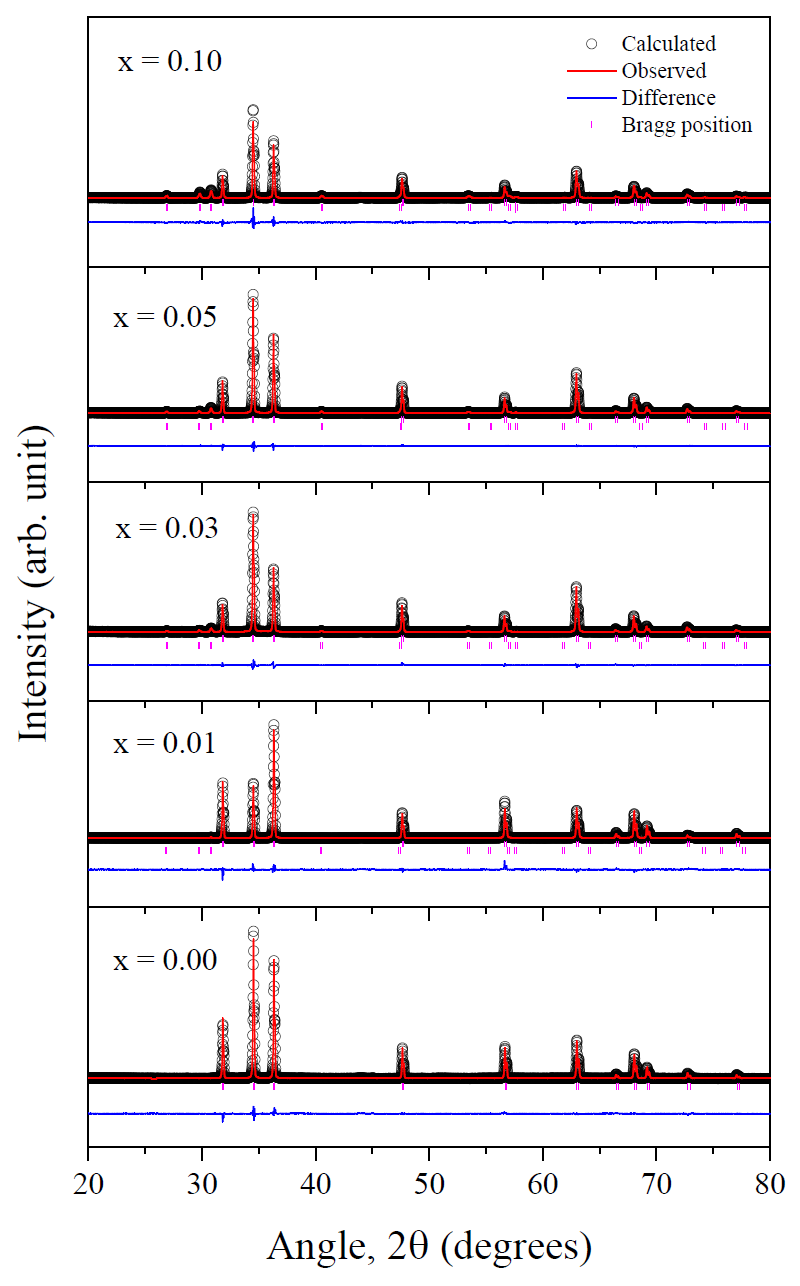}
\caption{Refined X-ray diffraction profiles of (Nd\textsubscript{0.5}Li\textsubscript{0.5})\textsubscript{x}Zn\textsubscript{1-x}O ceramics sintered at 1623 K.\label{fig:Fig. 2}}
\end{figure}

For further quantitative analysis of the structural parameters Rietveld refinement of the observed X-ray diffraction patterns have been carried out by using FULLPROF package \cite{26}. The refined XRD patterns of the (Nd\textsubscript{0.5}Li\textsubscript{0.5})\textsubscript{x}Zn\textsubscript{1-x}O bulk ceramics sintered at 1623 K are illustrated in Fig. \ref{fig:Fig. 2}.
It is evident from Fig. \ref{fig:Fig. 2} that the calculated patterns show rather good agreement with the experimental XRD patterns. Negligible difference between the experimental and calculated profiles is observed. The refined crystallographic parameters of (Nd\textsubscript{0.5}Li\textsubscript{0.5})\textsubscript{x}Zn\textsubscript{1-x}O ceramics are tabulated in Table \ref{table 1} with the corresponding fitting parameters. The values of R-factors are low indicating rather good refinement of the observed XRD profiles. The fractional coordinates of the studied compositions are listed in Table \ref{table 2}. The major phase of the co-doped samples are observed to be hexagonal structured ZnO with space group \textit{P6\textsubscript{3}mc} (186). The obtained lattice parameters are in good agreement with the other previous studies \cite{25, 27, 28}. However, all the compositions contain an impurity phase except the pristine ZnO. This extra phase is identified as hexagonal structured Nd\textsubscript{2}O\textsubscript{3} with space group \textit{P6\textsubscript{3}/mmc} (194) having lattice parameters nearly \textit{a} = 0.38317 nm, \textit{c} = 0.60089 nm and V = 0.07640 nm\textsuperscript{3}. In 1\% (Li, Nd) co-doped ZnO, the phase fractions are calculated to be 99.52\% ZnO and 0.48\% Nd\textsubscript{2}O\textsubscript{3}, respectively.   
\begin{table*}[ht]
\caption{\label{table 1}Refined crystallographic data of
(Nd\textsubscript{0.5}Li\textsubscript{0.5})\textsubscript{x}Zn\textsubscript{1-x}O ceramics.}
\begin{ruledtabular}
\begin{tabular}{cccccc}
\textbf{Parameters } & \textbf{x = 0.00} & \textbf{x = 0.01} & \textbf{x
= 0.03} & \textbf{x = 0.05} & \textbf{x = 0.10}\\
\hline
Crystal system &  &  & hexagonal &  &\\
Space group &  &  & \textit{P6\textsubscript{3}mc} (186) &  &\\
\emph{a} (nm) & 0.32457 & 0.32465 & 0.32487 & 0.32484 & 0.32474\\
\emph{c} (nm) & 0.51940 & 0.51959 & 0.51982 & 0.51986 & 0.51972\\
\emph{c/a} & 1.6002 & 1.6004 & 1.60008 & 1.6003 & 1.6004\\
\emph{\(\alpha\)}, \emph{\(\beta\)}, \emph{\(\gamma\)} (deg.) &  &  & 90,
90, 120 &  & \\
\emph{V} (nm\textsuperscript{3}) & 0.04738 & 0.04742 & 0.04751 & 0.04750
& 0.04746\\
\emph{Z } & 2 & 2 & 2 & 2 & 2\\
\emph{\(\chi\)}\textsuperscript{2} & 4.34 & 5.31 & 2.88 & 2.91 & 5.12\\
\emph{R\textsubscript{p}} (\%) & 5.83 & 8.34 & 6.42 & 6.00 & 8.85\\
\emph{R\textsubscript{wp }}(\%) & 8.07 & 11.0 & 8.42 & 8.28 & 11.5\\
\end{tabular}
\end{ruledtabular}
\end{table*}
\begin{table*}[ht]
\caption{\label{table 2}Fractional atomic coordinates of
(Nd\textsubscript{0.5}Li\textsubscript{0.5})\textsubscript{x}Zn\textsubscript{1-x}O ceramics.}
\begin{ruledtabular}
\begin{tabular}{ccccc}
\textbf{Composition } & \textbf{Atom} & \emph{\textbf{x}} &
\emph{\textbf{y}} & \emph{\textbf{z}}\\
\hline
x = 0.00 & Zn & 0.33333 & 0.66667 & 0.0\\
& O & 0.33333 & 0.66667 & 0.39971\\
x = 0.01 & Zn & 0.33333 & 0.66667 & 0.0\\
& O & 0.33333 & 0.66667 & 0.39237\\
x = 0.03 & Zn & 0.33333 & 0.66667 & 0.0\\
& O & 0.33333 & 0.66667 & 0.38532\\
x = 0.05 & Zn & 0.33333 & 0.66667 & 0.0\\
& O & 0.33333 & 0.66667 & 0.38313\\
x = 0.10 & Zn & 0.33333 & 0.66667 & 0.0\\
& O & 0.33333 & 0.66667 & 0.39381\\
\end{tabular}
\end{ruledtabular}
\end{table*}

\begin{figure}[ht]
\includegraphics[width=8cm]{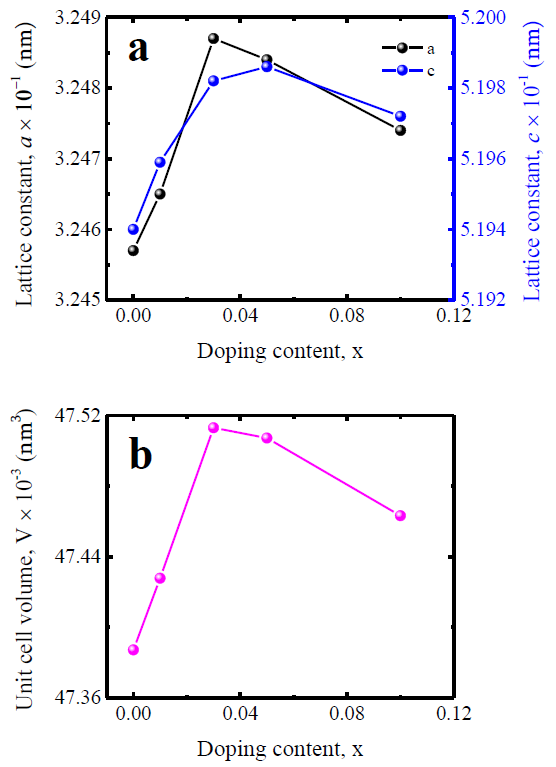}
\caption{(a) Variation of lattice constants and (b) unit cell volume, as a function of doping concentration of (Nd\textsubscript{0.5}Li\textsubscript{0.5})\textsubscript{x}Zn\textsubscript{1-x}O ceramics.\label{fig:Fig. 3}}
\end{figure}

Fig. \ref{fig:Fig. 3} illustrates the lattice parameters as a function of doping concentration (x). With the increase in doping content the lattice parameters are found to extend up to 3\% doping concentration. From 5\% doping concentration the lattice parameters begin to decrease. This result can be well explained from the perspective of ionic radius of the constituent ions. According to the valance balance one Nd\textsuperscript{3+} (0.098 nm) and one Li\textsuperscript{+} (0.068 nm) may substitute the 2Zn\textsuperscript{2+} (0.074 nm) ions. Hence the average ionic radius of the dopant ions is nearly 0.083 nm that is comparatively larger than the ionic radius of Zn\textsuperscript{2+}. Due to this reason the lattice parameters are found to extend up to 3\% doping concentration. When the doping content exceeds 3\%, Nd\textsuperscript{3+} dopant is not incorporated into the lattice site and may appear in the form of Nd\textsubscript{2}O\textsubscript{3} (resulting increased number of impurity peaks in the XRD pattern). However, due to the smaller ionic radius of Li\textsuperscript{+}, it can still replace Zn\textsuperscript{2+}. Due to this reason the lattice parameters are found to decrease from 5\% doping concentration.

\begin{figure}[ht]
\includegraphics[width=7cm]{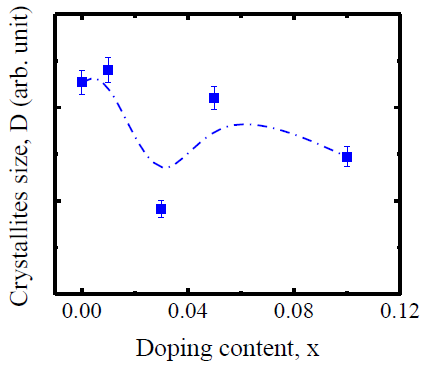}
\caption{Variation of crystallites size as a function of doping content of (Nd\textsubscript{0.5}Li\textsubscript{0.5})\textsubscript{x}Zn\textsubscript{1-x}O ceramics.\label{fig:Fig. 4}}
\end{figure}

The average crystallites size (\textit{D}) of the studied compositions are evaluated by using the well known Scherrer equation \cite{29, 30}, \textit{D} = (0.9\(\lambda\))/(\(\beta\)cos\(\theta\)), where \(\lambda\) is the wavelength of the X-rays, \(\beta\) is the width (full-width at half-maximum) of the X-ray diffraction peak in radians and \(\theta\) is the Bragg angle in degrees. Scherrer's formula is basically an approximation and used to estimate the average particles or crystallites size in the form of powder. This equation is only applicable for nano-scale range particles (100-200 nm) and basically inapplicable for grains larger than 0.1 to 0.2 \(\mu\)m \cite{30, 31}. However, though this equation does not provide accurate quantitative estimation for larger grains, it can provide good qualitative assessment. The evaluated crystallites size of (Nd\textsubscript{0.5}Li\textsubscript{0.5})\textsubscript{x}Zn\textsubscript{1-x}O ceramics is plotted in Fig. \ref{fig:Fig. 4} as a function of doping concentration. It is evident that on average the crystallites size is decreased with the increase in doping content. Some anomaly occurs due to the presence of secondary phase in the desired compositions.  

\subsection*{B. Density and Porosity}
\begin{figure}[ht]
\includegraphics[width=8cm]{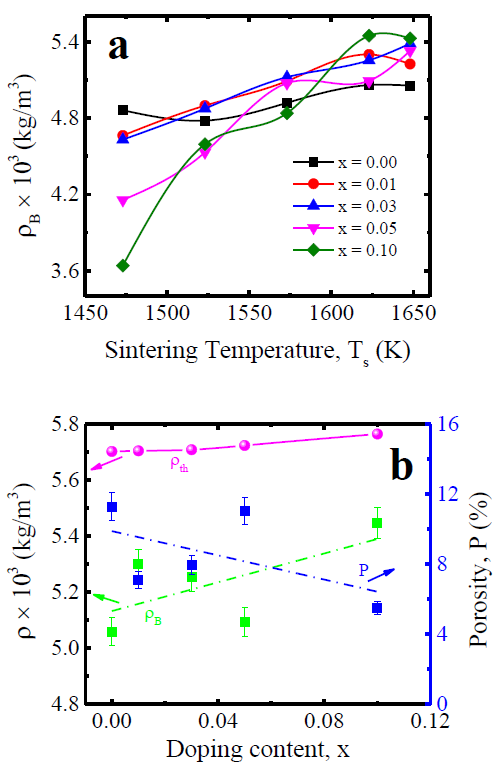}
\caption{(a) Sintering temperature dependent bulk density of (Nd\textsubscript{0.5}Li\textsubscript{0.5})\textsubscript{x}Zn\textsubscript{1-x}O ceramics. (b) Doping content dependent bulk density, theoretical density and porosity of (Nd\textsubscript{0.5}Li\textsubscript{0.5})\textsubscript{x}Zn\textsubscript{1-x}O ceramics sintered at 1623 K.\label{fig:Fig. 5}}
\end{figure}

In polycrystalline materials density plays a crucial role in controlling the dielectric and magnetic properties. Sintering temperature dependent bulk density (\(\rho\)\textsubscript{B}) is illustrated in Fig. \ref{fig:Fig. 5}(a). It is found that \(\rho\)\textsubscript{B} is enhanced up to 1623 K and begins to decrease after 1623 K. Generally, grain boundaries are grown over the pores by the force generated through the thermal energy during the sintering process. As a result, pore volume is decreased and the samples become more dense with the increase in T\textsubscript{s}. Porosity in ceramic sample is generated basically from two sources, intergranular porosity and intragranular porosity \cite{32}. The intergranular porosity depends upon the average grain size. However, some pores are trapped within the ceramic sample due to the very high driving energy at higher sintering temperature. Consequently, \(\rho\)\textsubscript{B} begins to decrease at very high T\textsubscript{s}.  

The values of doping content dependent bulk density, theoretical density and porosity are listed in Table \ref{table 3}. The bulk density, theoretical density and porosity as a function of doping concentration of (Nd\textsubscript{0.5}Li\textsubscript{0.5})\textsubscript{x}Zn\textsubscript{1-x}O ceramics sintered at 1623 K are also plotted in Fig. \ref{fig:Fig. 5}(b). It is observed that \(\rho\)\textsubscript{B} and \(\rho\)\textsubscript{th} is enhanced with doping content. This phenomenon can be explained in terms of atomic weight of the constituent atoms. The average atomic weight of Li (6.94 amu) and Nd (144.242 amu) is 75.591 amu that is larger than the atomic weight of Zn (65.38 amu). Since Zn is substituted by Li and Nd, it is expected to enhance the density of the studied compositions with the increase in doping content. Diffusion of oxygen ions through the ceramics during the sintering process can also accelerate the densification of the ceramic material. Some deviations occur from the linearly fitted line due to the presence of excess Nd\textsubscript{2}O\textsubscript{3} content that leads to inhomogeneities in the particles. The reason can also be attributed to the very high sintering temperature and sintering time. Since the melting temperature of Li (453.5 K) is very low than the sintering temperature there is possibility for enhancing the sinterability through liquid phase formation \cite{35}. On the other hand, porosity exhibits completely opposite trend. Porosity is decreased with the increase in doping content as the samples become more dense. 
\begin{table}[ht]
\caption{\label{table 3}Bulk density, theoretical density and porosity of
(Nd\textsubscript{0.5}Li\textsubscript{0.5})\textsubscript{x}Zn\textsubscript{1-x}O ceramics sintered at 1623 K.}
\begin{ruledtabular}
\begin{tabular}{cccc}
\textbf{x} & \textbf{\(\rho\)\textsubscript{B}} & \textbf{\(\rho\)\textsubscript{th}} & \textbf{P (\%)}\\
& \textbf{\(\times 10^3\) (kg/m\textsuperscript{3})} & \textbf{\(\times 10^3\) (kg/m\textsuperscript{3})} &\\ 
\hline
0.00 & 5.05 & 5.7019 & 11\\
0.01 & 5.29 & 5.7041 & 7\\
0.03 & 5.25 & 5.7082 & 8\\
0.05 & 5.09 & 5.7231 & 11\\
0.10 & 5.44 & 5.7642 & 6\\
\end{tabular}
\end{ruledtabular}
\end{table}

\subsection*{C. Morphological Analysis}
\begin{figure}[ht]
\includegraphics[width=8cm]{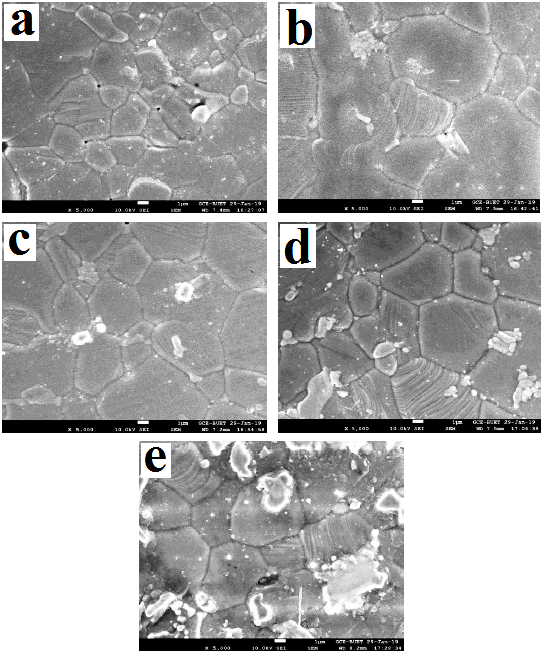}
\caption{FESEM micrographs of (Nd\textsubscript{0.5}Li\textsubscript{0.5})\textsubscript{x}Zn\textsubscript{1-x}O ceramics sintered at 1623 K. (a) x = 0.00, (b) x = 0.01, (c) x = 0.03, (d) x = 0.05 and (e) x = 0.10.\label{fig:Fig. 6}}
\end{figure}

The FESEM micrographs of (Nd\textsubscript{0.5}Li\textsubscript{0.5})\textsubscript{x}Zn\textsubscript{1-x}O ceramics sintered at 1623 K are demonstrated in Fig. \ref{fig:Fig. 6}. It is evident that substitution of Li and Nd at the Zn site of ZnO ceramics have a significant effect on the grain size. All the compositions contain randomly aligned non-uniform grains in size and shape due to very high sintering temperature. The distribution of grains is not homogeneous and some agglomeration is appeared in higher doped samples. Non-equivalent size of dopant ions may be the reason behind this agglomeration of grains.

\begin{figure}[ht]
\includegraphics[width=8cm]{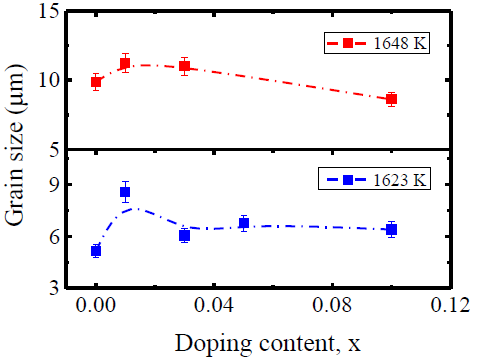}
\caption{Variation of the average grain size as a function of doping content for (Nd\textsubscript{0.5}Li\textsubscript{0.5})\textsubscript{x}Zn\textsubscript{1-x}O ceramics sintered at 1623K and 1648 K.\label{fig:Fig. 7}}
\end{figure}

The average grain sizes measured by using the linear intercept technique of the studied compositions as a function of doping content are demonstrated in Fig. \ref{fig:Fig. 7}. Since the melting temperature of Li (453.5 K) is comparatively low, it leads to enormous grain growth of zinc oxide with a little amount of Nd doping. This is the primary reason behind the enhancement of grain size with 1\% doping content. However, the grain size is decreased with further increase in dopant percentage due to the formation of large volume of Nd\textsubscript{2}O\textsubscript{3} secondary phase that interrupt the grain growth by pinning the grain boundaries and blocking mass transportation \cite{36}. However, It is observed that the average grain size is on average decreased with the increase in doping content showing similar trend as the variation of crystallites size as a function of doping content as shown in Fig. \ref{fig:Fig. 4}. This result accord well with others literature of donor acceptor co-doped ZnO ceramics \cite{33, 34}. The decrease in grain size with doping content can be attributed to the slower diffusion of larger Nd cation into ZnO ceramics. The increase in Li and Nd into ZnO leads to increase in the internal stress, resulting smaller grain size.

\subsection*{D. Energy Dispersive X-ray Analysis}
\begin{figure}[ht]
\includegraphics[width=8.0cm]{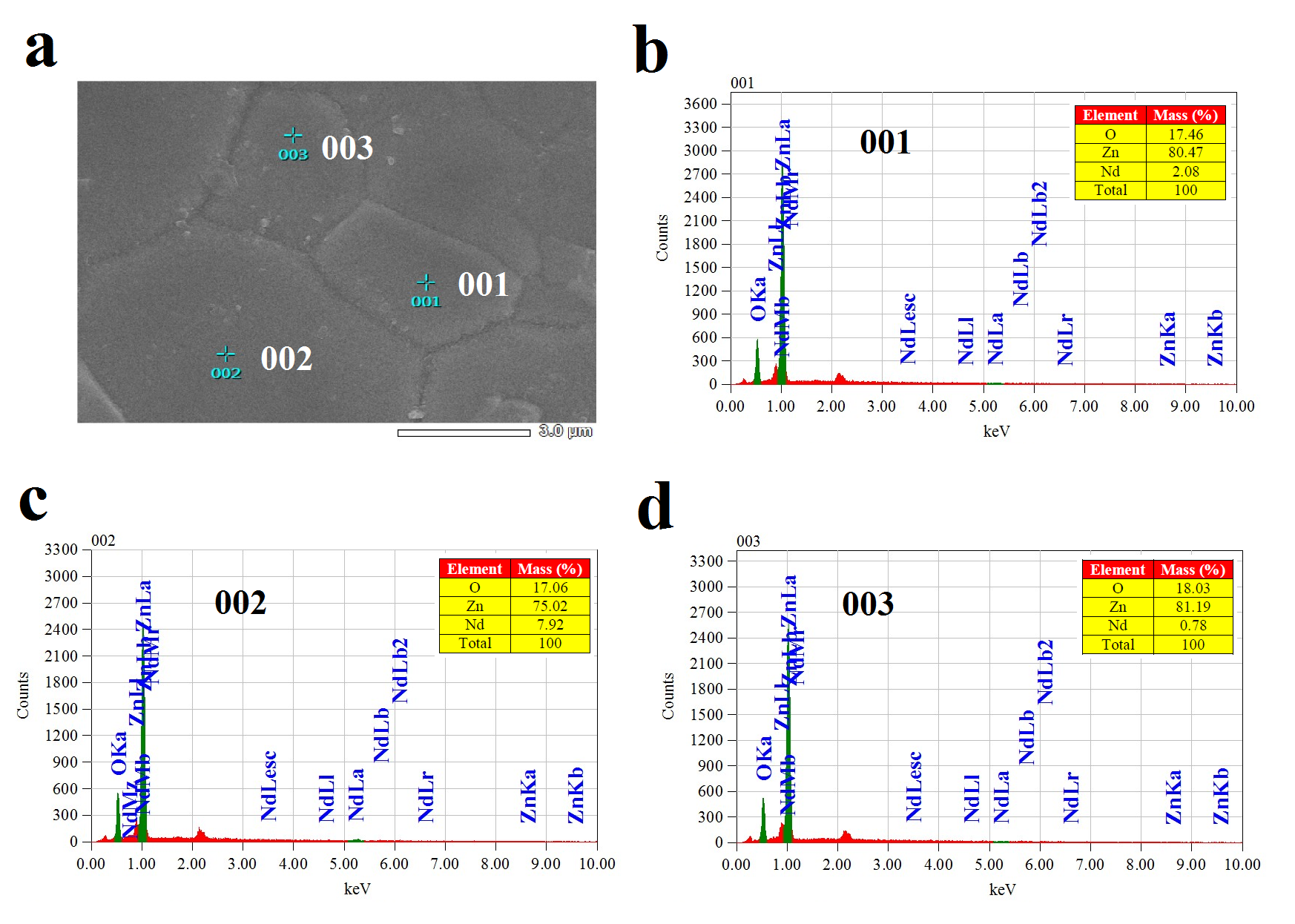}
\caption{EDX spectra of 1\% (Li, Nd) co-doped ZnO ceramics sintered at 1623 K. (a) Image used for EDX analysis. Information corresponding to (b) point 001, (c) point 002 and (d) point 003, of Nd\textsubscript{0.005}Li\textsubscript{0.005}Zn\textsubscript{0.99}O ceramics.\label{fig:Fig. 8}}
\end{figure}

For quantitative elemental analysis EDX spectra taken at various points of the sample is demonstrated in Fig. \ref{fig:Fig. 8}. The identified peaks are generated from O, Zn and Nd of the (Nd\textsubscript{0.5}Li\textsubscript{0.5})\textsubscript{x}Zn\textsubscript{1-x}O ceramics. However, peak corresponds to Li is missing since Li is a very light element and remains undetected in EDX analysis. It is found from the EDX spectra that there exist a well consistency between the mass percentage of the elements in the component phase and nominal composition of the corresponding phase. Some variation occurs due to very high calcination and sintering temperature.

\subsection*{E. Dielectric Properties}
\begin{figure}[ht]
\includegraphics[width=7.5cm]{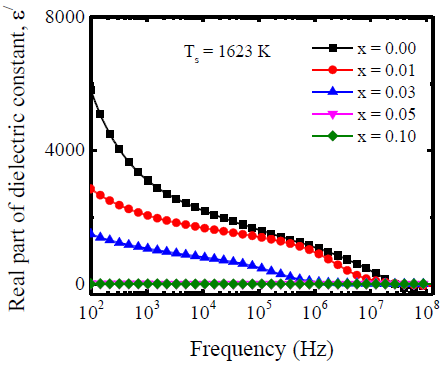}
\caption{Variation of dielectric constant as a function of frequency of (Nd\textsubscript{0.5}Li\textsubscript{0.5})\textsubscript{x}Zn\textsubscript{1-x}O ceramics sintered at 1623 K.\label{fig:Fig. 9}}
\end{figure}

Frequency dependent real part of dielectric constant (\(\varepsilon^\prime\)) of various (Nd\textsubscript{0.5}Li\textsubscript{0.5})\textsubscript{x}Zn\textsubscript{1-x}O ceramics sintered at 1623 K are illustrated in Fig. \ref{fig:Fig. 9}. 
It is evident that at lower frequency all the compositions exhibit higher values of dielectric constant. The permittivity is gradually decreased with the increase in frequency and becomes zero at very high frequency. Such frequency dependent dielectric behavior can be explained by Koops theory dependent on the Maxwell-Wagner model for inhomogeneous crystal structure \cite{37, 38, 39}. This model suggests that a typical dielectric medium is composed of well conducting grains (as shown in Fig. \ref{fig:Fig. 6}) that are generally separated by resistive (poorly conducting) grain boundaries. When an external electric field is applied on a dielectric medium the charge carriers begin to migrate through the conducting grain and are piled up at the resistive grain boundaries. As a result, large polarization (space-charge polarization) takes place within the dielectric medium resulting high dielectric constant. In this case the poorly conducting grain boundaries contribute to the higher value of permittivity at lower frequency. Different types of polarization mechanism is also responsible for the higher values of permittivity at low frequency region. In the low frequency region all of the four polarization mechanisms (ionic, electronic, dipolar and space-charge polarization) contribute to the total polarization in the compound resulting high dielectric constant. But with the increase in frequency (i.e., at high frequency region) the contribution of some of the above-mentioned polarization mechanism to the total polarization is terminated resulting lower values of permittivity.
\begin{figure}[ht]
\includegraphics[width=7.5cm]{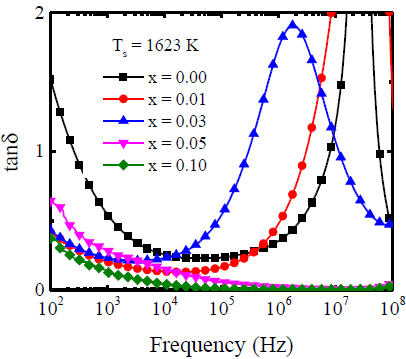}
\caption{Variation of loss tangent as a function of frequency of polycrystalline (Nd\textsubscript{0.5}Li\textsubscript{0.5})\textsubscript{x}Zn\textsubscript{1-x}O ceramics sintered at 1623 K.\label{fig:Fig. 10}}
\end{figure}

The inherent dissipation of electromagnetic energy as a form of heat from a dielectric material is defined as the dielectric loss. In conduction loss, the energy dissipation is caused by flowing of charge through a material. On the other hand, in dielectric loss the dissipation of energy is caused by the movement of charges in an alternating electromagnetic field as polarisation switches direction. Dielectric loss is high at the resonance or relaxation frequencies of the polarisation mechanisms as the polarisation lags behind the applied field, causing an interaction between the field and the dielectric’s polarisation that results in heating \cite{40}. Materials having higher dielectric constants generally show higher dielectric loss. The frequency dependent dielectric loss factor (tan\(\delta\)) of polycrystalline ceramics (Nd\textsubscript{0.5}Li\textsubscript{0.5})\textsubscript{x}Zn\textsubscript{1-x}O sintered at 1623 K are illustrated in Fig. \ref{fig:Fig. 10}. It is observed that at low frequency region the loss tangent is maximum and with the increase in frequency tan\(\delta\) is gradually decreased. The domain wall motion is suppressed at higher frequency and magnetization is changed by rotation resulting lower loss at higher frequency. However, there also appears some broad relaxation peaks at high frequency region. Under the influence of externally applied field, the electric dipoles tend to align with the electric field. But the alignment is not happened instantaneously. The dipoles take some time to align with the field. This phenomenon is known as dipole relaxation. When the relaxation or resonance frequency is equal to the applied frequency, a maximum in loss tangent may be occurred.  
\begin{figure}[ht]
\includegraphics[width=7.5cm]{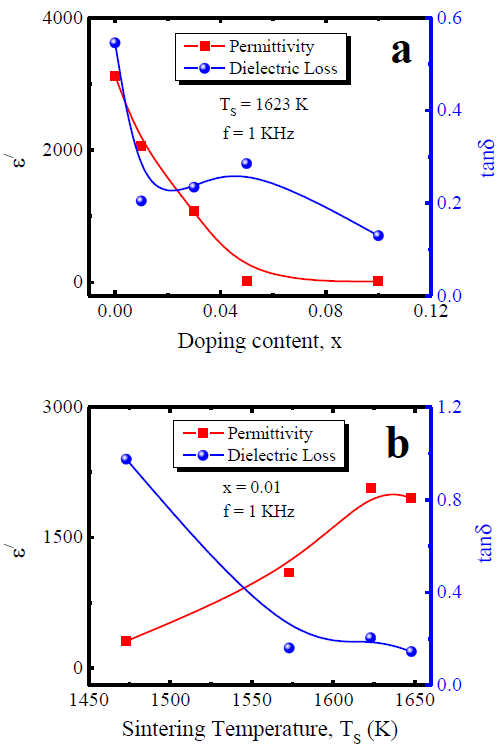}
\caption{(a) The permittivity and loss tangent as a function of doping concentration of (Nd\textsubscript{0.5}Li\textsubscript{0.5})\textsubscript{x}Zn\textsubscript{1-x}O ceramics sintered at 1623 K. (b) Sintering temperature dependent permittivity and loss tangent of 1\% (Li, Nd) co-doped ZnO ceramics.\label{fig:Fig. 11}}
\end{figure}

Fig. \ref{fig:Fig. 11}(a) illustrates the dependence of dielectric constant and dielectric loss factor on doping content of (Nd\textsubscript{0.5}Li\textsubscript{0.5})\textsubscript{x}Zn\textsubscript{1-x}O ceramics sintered at 1623 K. It is evident that the relative permittivity is decreased with the increase in doping concentration. This phenomenon can be explained by the grain size distribution in corresponding samples. The grain size is slightly extended [Fig. \ref{fig:Fig. 7}] for 1\% (Li, Nd) co-doping in ZnO, that allows to decrease the number of grains per unit volume resulting decrease the dipole moment of the whole system. As a result, the relative permittivity is decreased. However, for further increase in dopant content the grain size is decreased resulting further decrease in the relative permittivity. In general grains are made of multiple domains that are separated by domain walls. The value of dielectric constant depends upon the number of populations of domain and domain wall mobility \cite{41}. Since the grain size is relatively small and non-uniform throughout the sample, the movement of domain wall is relatively difficult and irregular. As a result, the value of relative permittivity is decreased. Moreover, it is also clear from Fig. \ref{fig:Fig. 11}(a) that the dielectric loss is also reduced with the increase in dopant concentration. Some anomaly occurs due to the presence of secondary phase in the desired samples. This phenomenon can be explained through the newly proposed dielectric polarization mechanism Electron Pinned Defect Dipole (EPDD) model \cite{42}. According to this model, electrons are created by the donor atom Nd\textsuperscript{3+}. These electrons are then localized by the presence of acceptor atom Li\textsuperscript{+}. These localized electrons lead to extra low dielectric loss \cite{43}.

Fig. \ref{fig:Fig. 11}(b) illustrates the dependence of dielectric constant and dielectric loss factor on sintering temperature of 1\% doped (Nd\textsubscript{0.5}Li\textsubscript{0.5})\textsubscript{x}Zn\textsubscript{1-x}O ceramics. The real part of dielectric constant is enhanced up to 1623 K and then begins to decrease for further increase in T\textsubscript{s}. It is well established that density plays a vital role to control the dielectric properties of ceramic materials. As shown in Fig. \ref{fig:Fig. 5}(a) the bulk density of 1\% doped composition is increased up to 1623 K and then begins to decrease for further increase in T\textsubscript{s} showing exactly the similar trend with the variation of permittivity as a function of T\textsubscript{s}. It is also observed that the dielectric loss is decreased with the increase in sintering temperature. However, It is clear from Fig. \ref{fig:Fig. 11} that 1\% (Li, Nd) co-doped ZnO shows slightly lower dielectric constant than pristine ZnO but relatively very low dielectric loss than pure ZnO ceramics sintered at 1623 K. Hence 1\% (Li, Nd) co-doped ZnO sintered at 1623 K can be regarded as the optimum sample.

\subsection*{F. The ac-conductivity}
\begin{figure}[ht]
\includegraphics[width=7.5cm]{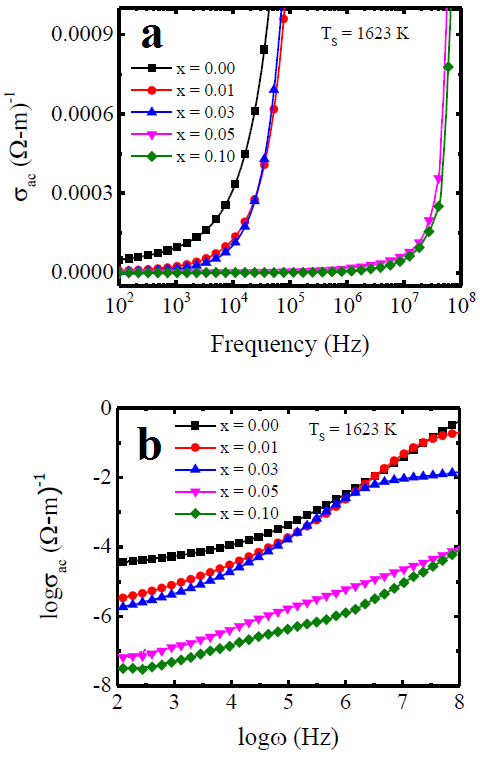}
\caption{(a) Variation of ac-conductivity with frequency and (b) Variation of log\(\sigma\)\textsubscript{ac} with log\(\omega\) of polycrystalline (Nd\textsubscript{0.5}Li\textsubscript{0.5})\textsubscript{x}Zn\textsubscript{1-x}O ceramics sintered at 1623 K.\label{fig:Fig. 12}}
\end{figure}

For comprehending the conduction mechanism in different materials ac-conductivity \(\sigma\)\textsubscript{ac} is a crucial parameter. The frequency dependent ac-conductivity of the studied compositions sintered at 1623 K is illustrated in Fig. \ref{fig:Fig. 12}(a). The value of \(\sigma\)\textsubscript{ac} of all the sintered samples is found almost constant at low frequency region. Whereas the value of \(\sigma\)\textsubscript{ac} is increased very fast after a certain frequency. In low frequency region, the conductivity is almost independent of frequency because in this region the resistive grain boundaries are very active according to the Maxwell-Wagner double layer model \cite{44, 45}. On the other hand in high frequency region (hopping region), the conductivity increases faster because of the very active conductive grains thereby enhances hopping of charge carriers that contributes to the rise in conductivity. Therefore, the transport phenomenon occurs through penetrating process in low frequency zone, whereas in high frequency zone  the transport phenomena is kept up by hopping carriers that are generated from the substituted elements \cite{46}.

The frequency dependent ac-conductivity can also be explained through the polaron hopping model proposed by Austin and Mott \cite{47}. Variation of log\(\sigma\)\textsubscript{ac} with log\(\omega\) of the studied compositions sintered at 1623 K is illustrated in Fig. \ref{fig:Fig. 12}(b). According to the small polaron hopping model (large polaron hopping model) the value of \(\sigma\)\textsubscript{ac} increases (decreases) with the enhancement of frequency \cite{48}. It is evident from Fig. \ref{fig:Fig. 12}(b) that the mechanism of conduction of the studied samples can be explained by the small polaron hopping model since the value of \(\sigma\)\textsubscript{ac} increases for all samples with the increase in frequency.
\begin{figure}[ht]
\includegraphics[width=7.5cm]{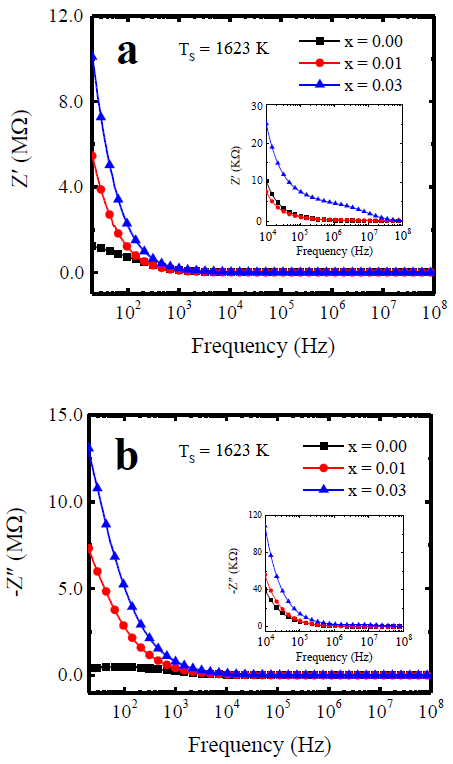}
\caption{Variation of (a) \(Z^\prime\) and (b) \(Z^{\prime\prime}\), as a function of frequency of (Nd\textsubscript{0.5}Li\textsubscript{0.5})\textsubscript{x}Zn\textsubscript{1-x}O ceramics sintered at 1623 K.\label{fig:Fig. 13}}
\end{figure}

\begin{figure}[ht]
\includegraphics[width=7.5cm]{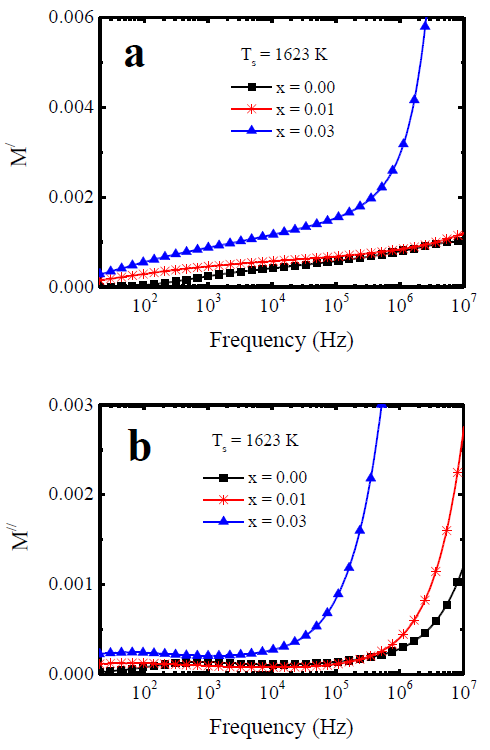}
\caption{Variation of (a) \(M^\prime\) and (b) \(M^{\prime\prime}\), as a function of frequency of polycrystalline (Nd\textsubscript{0.5}Li\textsubscript{0.5})\textsubscript{x}Zn\textsubscript{1-x}O ceramics sintered at 1623 K.\label{fig:Fig. 14}}
\end{figure}

\subsection*{G. Complex Impedance Spectra Analysis}
Complex impedance spectroscopy (CIS) is a very powerful approach for characterizing the electrical properties of materials. Fig. \ref{fig:Fig. 13} illustrates the frequency dependent real part \(Z^\prime\) and imaginary part \(Z^{\prime\prime}\) of the impedance for (Nd\textsubscript{0.5}Li\textsubscript{0.5})\textsubscript{x}Zn\textsubscript{1-x}O ceramics sintered at 1623 K. It is evident that the value of \(Z^\prime\) decreases sharply up to 1 KHz and then remains nearly constant at high frequency region implying the increase in electrical conductivity of the studied compositions. However, at high frequency region the plateau like nature of \(Z^\prime\) implies possible occurrence of space charge polarization under the influence of external applied field \cite{49}. The imaginary part of impedance \(Z^{\prime\prime}\) exhibits similar nature as \(Z^\prime\) [Fig. \ref{fig:Fig. 13}(b)]. No relaxation peak is appeared indicating the absence of immobile charges in (Nd\textsubscript{0.5}Li\textsubscript{0.5})\textsubscript{x}Zn\textsubscript{1-x}O ceramics.

Fig. \ref{fig:Fig. 14}(a) shows the frequency dependent real part of electric modulus \(M^\prime\) of (Nd\textsubscript{0.5}Li\textsubscript{0.5})\textsubscript{x}Zn\textsubscript{1-x}O ceramics sintered at 1623 K. At lower frequency the magnitude of \(M^\prime\) is nearly zero indicating the ease of polaron hopping \cite{50, 51}. It also implies that the contribution of electrode effect in total impedance is very negligible in case of the studied compositions. However, the magnitude of \(M^\prime\) increases gradually with the increase in frequency for all the compositions. The frequency dependent imaginary part of dielectric modulus \(M^{\prime\prime}\) is shown in Fig. \ref{fig:Fig. 14}(b). No relaxation peak is observed. Therefore, charge carriers are mobile in between the grains over the whole frequency range studied in the present case \cite{52}. However, with the increase in frequency the magnitude of \(M^{\prime\prime}\) in enhanced. Hence, at higher frequency the charge carriers may be trapped in a potential well and are allowed to mobile over short distances (inside the grains) \cite{52}. 
\begin{figure}[ht]
\includegraphics[width=7.5cm]{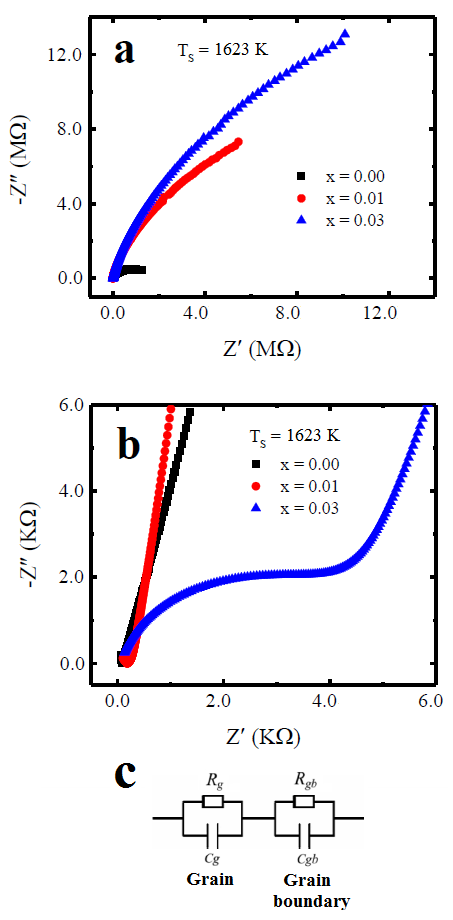}
\caption{The Cole-Cole plot of polycrystalline (Nd\textsubscript{0.5}Li\textsubscript{0.5})\textsubscript{x}Zn\textsubscript{1-x}O ceramics sintered at 1623 K. (a) \(Z^{\prime\prime}\) vs \(Z^\prime\) plot at low frequency region. (b) \(Z^{\prime\prime}\) vs \(Z^\prime\) plot at high frequency region. (c) Equivalent circuit model.\label{fig:Fig. 15}}
\end{figure}

According to the brick-layer model \cite{53} a polycrystalline ceramics can be represented by an equivalent circuit composed of three parallel RC components. By using these phenomena it is possible to separate the specific contribution of grain and grain boundaries to the total resistance. These three parallel RC circuits are connected in series combination with one another that corresponds to the grains, grain boundaries and electrode effect. In a Nyquist plot (\(Z^{\prime\prime}\) vs \(Z^\prime\) plot) each RC component of the equivalent circuit generates a semicircle. Formation of a single semicircle indicates only the grain effects in the total resistance. Similarly, the formation of second and third semicircular arc corresponds to the grain boundary and electrode effects, respectively. Fig. \ref{fig:Fig. 15} demonstrates the Cole-Cole plot of (Nd\textsubscript{0.5}Li\textsubscript{0.5})\textsubscript{x}Zn\textsubscript{1-x}O ceramics sintered at 1623 K. It is observed that depressed semicircular arcs are formed indicating non Debye-type relaxation in (Nd\textsubscript{0.5}Li\textsubscript{0.5})\textsubscript{x}Zn\textsubscript{1-x}O ceramics. Single semicircular arcs are formed for all the samples except 3\% doped compositions. Two semicircular arcs are found in case of 3\% doped ZnO ceramics. Therefore, dielectric contribution comes from both the grain and grain boundary resistance in case of 3\% doped profile since semicircle at high frequency [Fig. \ref{fig:Fig. 15}(b)] indicates only the grain effect (bulk resistance) and semicircle at low frequency indicates grain boundary effect to the total resistance. However, pristine and 1\% doped profile possesses only single semicircle [Fig. \ref{fig:Fig. 15}(a)] which implies that in these compositions the major part of dielectric contribution comes from the grain boundary resistance. The dielectric properties of (Nd\textsubscript{0.5}Li\textsubscript{0.5})\textsubscript{x}Zn\textsubscript{1-x}O ceramics can be represented by an equivalent RC circuit depicted in Fig. \ref{fig:Fig. 15}(c).

\subsection*{H. Theoretical Analysis}
\begin{figure}[ht]
\includegraphics[width=8cm]{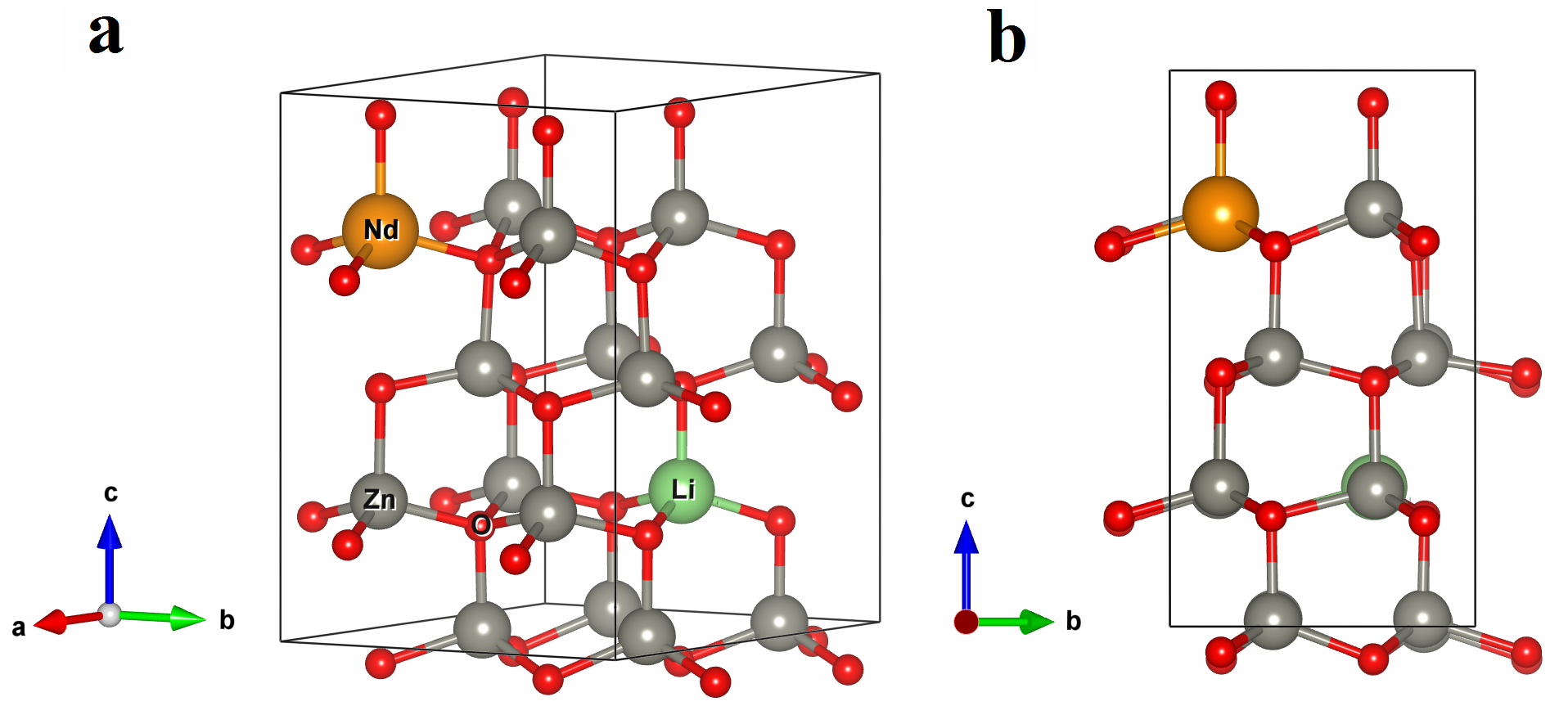}
\caption{The crystal structure (2 × 2 × 2 supercell) of (Li, Nd) co-doped ZnO ceramics. (a) Three dimensional and (b) two dimensional view.\label{fig:Fig. 16}}
\end{figure}

For simulating the doping effect in hexagonal wurtzite ZnO, 2 × 2 × 2 supercell (8 times the unit cell of the pure ZnO) is constructed. The supercell of ZnO contains 32 atoms including 16 Zn atoms and 16 O atoms. For inserting the impurities in pristine ZnO, two Zn atoms are replaced by Li and Nd atoms (substitutional doping) as shown in Fig. \ref{fig:Fig. 16} that produces a doping concentration of 12.5 atom\%. Hence the new formula of the co-doped ZnO can be written as (Nd\textsubscript{0.5}Li\textsubscript{0.5})\textsubscript{x}Zn\textsubscript{1-x}O (x = 0.125).

\begin{figure}[ht]
\includegraphics[width=7.5cm]{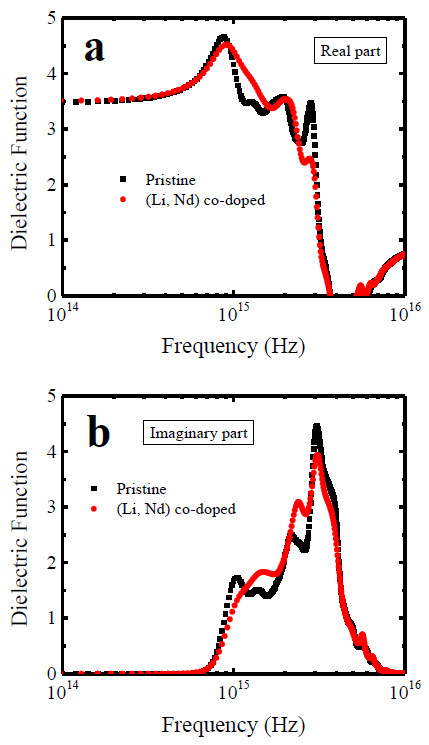}
\caption{The frequency dependent theoretically evaluated (a) real and (b) imaginary part of dielectric function for both the pure and (Li, Nd) co-doped ZnO ceramics.\label{fig:Fig. 17}}
\end{figure}

The real and imaginary part of dielectric function evaluated by using the DFT based theoretical approach for both the pristine and (Li, Nd) co-doped ZnO is shown in Fig. \ref{fig:Fig. 17}. It is evident that the value of real part of dielectric constant [Fig. \ref{fig:Fig. 17}(a)] is decreased with the increase in frequency for both the pure and co-doped profile showing concordance with the experimental observation [Fig. \ref{fig:Fig. 9}]. It should be noted that the theoretically evaluated dielectric function is for very high frequency (UV-VIS range). Hence, only the ionic and electronic polarization contribute to the total permittivity value. However, co-doping of Li and Nd into the ZnO ceramics does not have any major effect on the dielectric constant of pure ZnO. As shown in Fig. \ref{fig:Fig. 17}(b) the imaginary part of dielectric function of the co-doped sample is slightly less than that of the pristine ZnO. It implies that the ohmic loss is slightly reduced after co-doping showing good concurrence with experimental observation [Fig. \ref{fig:Fig. 10}].          

\section*{IV. CONCLUSIONS}

In summary, various (Nd\textsubscript{0.5}Li\textsubscript{0.5})\textsubscript{x}Zn\textsubscript{1-x}O ceramics (x = 0.00, 0.01, 0.03, 0.05, and 0.10) have been prepared and examined their structural, morphological and dielectric properties through different experimental techniques. Structural refinement confirms the formation of wurtzite hexagonal structured zinc oxide. In higher doped ZnO a secondary phase of Nd\textsubscript{2}O\textsubscript{3} is appeared due to the larger ionic radius of Nd\textsuperscript{3+} compared to Zn\textsuperscript{2+}. FESEM micrographs ensure that all the compositions contain randomly aligned non-uniform grains in size and shape due to very high sintering temperature. The distribution of grains is not homogeneous and some agglomeration is appeared in higher doped samples due to non-equivalent size of dopant ions. The value of dielectric constant is decreased with the increase in doping content. However, 1\% (Li, Nd) co-doped ZnO shows slightly lower dielectric constant (\(\approx\) 2066 at 1 KHz) than pristine ZnO but relatively very low dielectric loss (\(\approx\) 0.20 at 1 KHz) at room temperature than pure ZnO ceramics sintered at 1623 K. The grain size distribution and density of the studied compositions are two tuning factors to control the dielectric properties of these ceramics. The mechanism of ac-conductivity in the studied ceramics can be explained by a small polaron hopping model. Complex impedance study suggests that the dielectric relaxation mechanism in these ceramics is generally non Debye-type. Finally, the DFT study of (Li, Nd) co-doped ZnO suggests that before experimental fabrication DFT simulation can be used as an effective predictor of dielectric properties of the desired sample. The results presented here thus provide an insight of tuning the dielectric properties of various metal oxides.

\section*{ACKNOWLEDGMENTS}

This work is supported by the CASR of Bangladesh University of Engineering and Technology, Dhaka, Bangladesh [Grant No. 312(15)].

\renewcommand{\section}[2]{}%

\end{document}